\documentclass[english,twocolumn,prb,showpacs]{revtex4}
\usepackage[T1]{fontenc}
\usepackage[latin9]{inputenc}
\usepackage{graphicx}
\usepackage{amsmath}
\def\beq{\begin{equation}}
\def\eeq{\end{equation}}
\def\beqarr{\begin{eqnarray}}
\def\eeqarr{\end{eqnarray}}

\makeatletter

\usepackage{babel}
\makeatother
\begin{document}

\title{A Convenient 
Alternative for Series Manipulation via the Translation Operator}
\author{D. J. Priour, Jr$^{1}$}
\affiliation{1. Department of Physics,
University of Missouri, Kansas City, MO 64110-2499}

\date{\today}

\begin{abstract}
We derive and discuss a technique for manipulating power series
which is complementary to standard procedures.  We begin with 
the translation operator, but we express the operator as an infinite 
product instead of expanding it as a series and  
we apply combinatorial arguments to generate the 
terms in the series in an efficient manner 
with a minimum of clutter and intermediate calculations.
The method is effective 
for developing multivariate expansions, and may also be used to
manipulate series, e.g. in operations where one must take the 
reciprocal of a power series or raise it to a power that may be
fractional or irrational.  
In the case of two component perturbations, we obtain analytic 
expressions for the expansion coefficients.  We use
our technique to generate an electrostatic multipole expansion as a
demonstration of its utility in producing coefficients of special
functions such as the Legendre and Hermite polynomials.
\end{abstract}
 
%\pacs{}
%\vspace{-.5cm}

\maketitle

\section{Introduction}

Taylor series expansions are an important tool in the analysis of the 
effect of small perturbations where the development of a power series 
in the perturbing influence conveniently organizes successive corrections
to the zeroth order result.  If a single variable function is written 
as $f(x_{0} + \Delta x)$, where $\Delta x$ is the perturbation of small
displacement, then the Taylor has the form
\begin{equation}
f(x_{0} + \Delta x) = f(x_{0}) + \Delta x \frac{d}{dx}f(x_{0}) + \frac{1}{2!}
\Delta x^{2} \frac{d^{2}}{dx^{2}}f(x_{0})  + \ldots 
\label{eq:eq30}
\end{equation}
A salient example familiar to the modern physics student is the 
case of the Lorentz factor central to formulae from special relativity.
The Lorentz factor appears in the expression for the total energy of 
a body in motion, $\gamma m_{0}c^{2}$, where $\gamma = (1 - \beta^{2})^{-1/2}$,
$m_{0}$ is the rest mass, and $c$ is the speed of light; $\beta = v/c$ is the 
object's speed relative to the speed of light.
With a single variable Taylor series expansion, we readily obtain the 
energy of the mass when it is at rest, the classical kinetic energy term,
and the first relativistic correction.  One has
\begin{equation}
E = \gamma m_{0} c^{2} = (1 - \beta^{2})^{-1/2} m_{0} c^{2}
\end{equation}
For ordinary speeds (i.e. for very small $\beta$), the Lorentz factor
can be expanded as 
\begin{equation}
\gamma = (1 - \beta^{2})^{-1/2} = 1 + \frac{1}{2} \beta^{2} + 
\frac{3}{4} \beta^{4} + \ldots
\end{equation}
Thus, we find for the total energy of the moving mass
\begin{gather}
E = \gamma m_{0} c^{2} = m_{0}c^{2} + \frac{1}{2} m_{0} v^{2} + 
\frac{3}{4} m_{0} v^{4}/c^{4}\\ = m_{0} c^{2} + \frac{1}{2} m_{0} v^{2}
\left( 1 + \frac{3}{2} \beta^{2} \right),
\end{gather}
with the kinetic energy being augmented by the factor $1+ \frac{3}{2} 
\beta^{2}$.The latter correction becomes more important at relativistic
speeds where the kinetic energy scales more rapidly than the Newtonian
quadratic dependence.

For situations involving more than one perturbing influence such 
as the function $f(x_{0}+\Delta x, y_{0} + \Delta y,z_{0} + \Delta z )$
where there are several displacements,
a multivariable Taylor series expansion is often appropriate.
As an example, one could have an expression raised to a fractional power
such as $(1 + [\Delta x + \Delta y + \Delta z])^{\alpha}$ where the
perturbations are combined in a single term; in fact we will see that
broadly speaking, the 
analysis of expressions of the form $f(1 + [\Delta x + \Delta y + \Delta z])$
will be facilitated with the techniques we report on here.
To develop the 
power series in the displacements, one could begin with
\begin{gather}
\nonumber
f[1 + (\Delta x + \Delta y + \Delta z)] = 
1 + f^{'}(1)[\Delta x + \Delta y + \Delta z ] \\ 
+ \frac{1}{2!}f^{''}(1)[\Delta x + \Delta y + \Delta z]^{2}
+ \ldots
\label{eq:eq7}
\end{gather}
From Eq.~\ref{eq:eq7}, we see that we have first the task of generating the 
primary series and then the evaluation of the individual terms, 
which will be polynomials (i.e. binomials for the case of
two variables, trinomials for the three variable case,  etc.)
raised to successively increasing integer powers.

The case of a perturbation which happens to consist of a 
power series expansion in a small
parameter can also be a computationally intensive scenario if one 
wishes to calculate a perturbative expansion in the small quantity.
In the discussion which follows, we will use $x$ as a generic label for
the latter.  As an example, one may consider $\sec (x) = 
1/ \cos (x)$, where 
\begin{gather}
\sec (x)  
= (1 - x^{2}/2! + x^{4}/4! - x^{6}/6! + \ldots)^{-1} \\
= 1 - (-x^{2}/2! + x^{4}/4! - x^{6}/6! + \ldots) + \\
(-x^{2}/2! + x^{4}/4! - x^{6}/6! + \ldots)^{2}
+ \ldots
\label{eq:eq8}
\end{gather}
Again, even after the initial expansion, 
the individual terms will need to be evaluated and like powers of $x$ 
combined.  

We now discuss a complementary and less laborious approach for the development
of a Taylor series where the perturbation has a composite form such as 
the multivariable case in Eq.~\ref{eq:eq7} 
or the situation where the perturbing 
influence is itself a power series in a smaller parameter, $x$.
Our technique is based on the translation operator which can 
induce a coordinate shift by generating a 
Taylor series in one or more variable, and the method 
eliminates intermediate calculations while minimizing clutter.  

\section{The single variable case}

We comment first on the 
single variable case; excellent discussions may be found in standard
texts~\cite{uno}. 
Consider a function $f(x_{0} + \Delta x)$ where as before there
is a shift about $x_{0}$ of $\Delta x$.
For infinitesimal translations $\delta x$, reasonably behaved
functions can be accurately rendered in linearized form with
$f(x_{0} + \delta x) = f(x_{0}) + \delta x f^{'}(x_{0})$.
In terms of the differential operator $\hat{d}/dx$ one has,
equivalently,
\begin{equation}
f(x_{0} + \delta x) = (1 + \delta x \hat{d}/dx) f(x)|_{x = x_{0}}
\label{eq:eq10}
\end{equation}
It is possible to build up several shifts with 
\begin{equation}
f(x_{0} + n \delta x) = (1 + \delta x \hat{d}/dx)^{n} f(x)|_{x = x_{0}} 
\label{eq:eq11}
\end{equation}
where there are $n$ copies of the differential translation operator 
$(1 + \delta \hat{d}/dx)$.  This logic suggests that we may more accurately
take into a finite shift $\Delta x$ by combining many smaller translations 
to set up the larger shift.  Hence, it is reasonable to suggest that  
\begin{equation}
f(x_{0} + \Delta x) = \lim_{N \rightarrow \infty} 
(1 + \Delta x/N)^{N} 
\end{equation}
becomes an increasingly accurate approximation
as $N$ becomes very large.
The identity $\displaystyle\lim_{N \rightarrow \infty} 
( 1 + \Delta x/N)^{N} = 
e^{\Delta x}$ suggests that we associate a 
finite translation $\Delta x$ with the 
action of the exponentiated differential operator and
well known result
\begin{equation}
\hat{T}(\Delta x) = e^{\Delta x \hat{d}/dx} \equiv
\lim_{N \rightarrow \infty} (1 + \Delta x/N)^{N}
\label{eq:eq13}
\end{equation}

In conjunction with this result, we apply combinatorial logic to develop 
a series in $\Delta x \hat{d}/dx$. Although the series will 
turn out to be nothing 
other than the standard single variable Taylor expansion
given in Eq~\ref{eq:eq30}, a very similar 
approach will be applied to obtain the terms of the multivariable 
expansion where there are more intricacies to navigate.
Returning to the infinite product form, $(1 + \Delta x/N 
\hat{d}/dx )^{N}$, we examine how the power series will take shape
as $N$ becomes very large.  

The zeroth order term in the power series 
will clearly be unity.  To obtain the first 
order term, we must mix in a factor of $\frac{\Delta x}{N} \frac{\hat{d}}{dx}$,
and there are a total of $N$ ways of doing this.  For the second order 
term, there are $N$ ways to choose the first factor of 
$\frac{\Delta x}{N} \frac{\hat{d}}{dx}$, and $N-1$ ways to choose the second
factor.  However, with the order not being relevant, 
there will be redundant terms, and we must 
divide by a permutation factor of $2!$ to compensate for double 
counting.  Proceeding in a similar way, we see that 
the prefactor for the third order term
will be $N(N-1)(N-2)/3!$. Combining these results, we see that the 
entire series will have the form
\begin{equation}
(1 + \frac{\Delta x}{N} )^{N} f(x) = 1 + \Delta x \frac{N}{N}\frac{df}{dx}
+ \Delta x^{2} \frac{N(N-1)}{2!N^{2}} \frac{d^{2} f}{dx^{2}} + \ldots
\label{eq:eq14}
\end{equation}
Finally, in the large $N$ limit, factors involving $N$ cancel and
we obtain the anticipated result
\begin{equation}
\left ( 1 + \frac{\Delta x}{N} \frac{\hat{d}}{dx} \right )^{N} = 
1 + \Delta x \frac{df}{dx} 
+ \frac{1}{2!} \Delta x^{2}\frac{d^{2}}{dx^{2}} + \dots
\label{eq:eq15}
\end{equation}

\section{The multivariable case}

For the multivariable case $f(y_{1} + \Delta_{1}, 
y_{2} + \Delta_{2}, \ldots)$, we will effect the translation with 
\begin{gather}
\hat{T}(\Delta_{1}, \Delta_{2}, \ldots , \Delta_{n}) = \\
\exp \left (\Delta_{1} \frac{\hat{\partial}}{\partial y_{1}}+ 
\Delta_{2} \frac{\hat{\partial}}{\partial y_{2}} + \ldots + 
\Delta_{n} \frac{\hat{\partial}}{\partial y_{n}} \right)  
\label{eq:eq16}
\end{gather}
Noting that $\Delta_{1} \frac{\hat{\partial}}{\partial y_{1}}+
\Delta_{2} \frac{\hat{\partial}}{\partial y_{2}} + \ldots +
\Delta_{n} \frac{\hat{\partial}}{\partial y_{n}}= 
\vec{\Delta y} \cdot \vec{\hat{\nabla}}$, this becomes
\begin{equation}
\hat{T}(\Delta_{1}, \Delta_{2}, \ldots , \Delta_{n}) = 
\exp \left( \vec{\Delta y} \cdot \vec{\hat{\nabla}} \right)
\end{equation}
En route to the multivariable Taylor series, one
might develop a power series in $\vec{\Delta y} \cdot \vec{\hat{\nabla}}$,
but the result will be a multivariable Taylor series requiring the 
subsequent evaluation of terms with polynomials raised to integer
powers as discussed earlier.
Instead, just as we did for the single variable case, we write 
the multivariable expansion as an infinite product in lieu of
an infinite series.  The result is 
\begin{gather}
\hat{T}(\Delta_{1}, \Delta_{2}, \ldots, \Delta_{n}) = \\
\left( 1 + \frac{1}{N} \left[ \Delta_{1} \frac{\hat{\partial}}
{\partial y_{1}} + \Delta_{2} \frac{\hat{\partial}}{\partial y_{2}}
+ \ldots  + \Delta_{n} \frac{\hat{\partial}}{\partial y_{n}}
\right] \right)^{N},
\label{eq:eq17}
\end{gather}
where it is understood that the large $N$ limit will be taken.

Using combinatorial logic in conjunction with Eq.~\ref{eq:eq17} will 
permit the immediate enumeration of specific terms.  As an 
example, the coefficient for the linear term $\Delta_{i}$ will
be $\Delta_{i} \frac{N}{N} \frac{\partial}{\partial y_{i}} f(y_{1},y_{2}, 
\ldots y_{n})$ with the $N$ factors canceling.  For quadratic terms,
there are two cases.  When $i \neq j$ there are $N(N-1)$ ways
of collecting terms with the factors $\Delta_{i} \Delta_{j}$; we 
obtain $\Delta_{i} \Delta_{j} \frac{\partial^{2}}{\partial y_{i}
\partial y_{j}} f(y_{1}, y_{2}, \ldots )$ in the large $N$ limit.
Note that since $i$ and $j$ are distinct, there is no need
to divide by a redundancy factor.
On the other hand, when $i = j$, we need to compensate for double counting
by dividing by $2!$, and the term is
 $\frac{1}{2!} \Delta_{i}^{2} \frac{\partial^{2}}{ 
\partial y_{i}^{2}} f(y_{1},y_{2}, \ldots y_{n})$ as $N \rightarrow \infty$.

In general, one obtains a partial derivative $\partial /\partial y_{j}$ 
for each copy of $\Delta_{j}$, and we must divide by
the permutation factor $n_{j}!$
to compensate for multiple counting. As an example, the coefficient of
the term $\Delta_{1} \Delta_{2}^{2} \Delta_{3} \Delta_{4}^{4}$
would be $\frac{1}{2!}\frac{1}{4!} \frac{\partial^{8}}{\partial y_{1} 
\partial y_{2}^{2} \partial y_{3} \partial y_{4}^{4}} 
f(y_{1}, y_{2}, y_{3}, y_{4},\ldots, y_{n})$.

A salient and frequently occurring case where we will gain an advantage 
are situations in which the perturbing elements are combined together
in a single term, i.e. as in $f(1 + [\Delta_{1} + 
\Delta_{2} + \ldots + \Delta_{n}])$. 
An important example is the case of an 
expression raised to a power $\alpha$ which needn't be an integer
(though there are nontrivial cases that are often of 
interest such as the reciprocal, where
$\alpha = -1$).
Our expression is hence
\begin{gather}
f(y_{1} + \Delta_{1}, y_{2} + \Delta_{2}, \ldots ,y_{n} 
+ \Delta_{n}) \\
= (1 + [\Delta_{1} + \Delta_{2} + \ldots
+ \Delta_{n} ] )^{\alpha}
\end{gather}
What will work in our favor and facilitate the calculation 
both in this special case and the broader class of 
functions $f(1 + [\Delta_{1} + \Delta_{2} + \ldots + 
\Delta_{n}])$
is the fact that 
the partial derivatives depend only on the total order of the 
derivative, not the variable with which one differentiates.
As an example, the derivatives $\frac{\partial^{3}}{\partial y_{1} 
\partial y_{2} \partial y_{3}}f(y_{1},y_{2},y_{3},\ldots,y_{n}) = 
\frac{\partial^{3}}{\partial y_{4}^{3}}f(y_{1},y_{2},y_{3},\ldots,
y_{n})
= \alpha (\alpha -1)(\alpha - 2)$ are identical.  Let us examine
the consequences of this simplifying property.

For linear terms, the partial derivative $\frac{\partial f}{\partial y_{i}}$
yields a factor $\alpha$ (independent of the value of $i$), 
and this value is the expansion coefficient.
For the quadratic terms, the partial differentiation will yield for
each $\left \{ i,j \right \}$ the same factor $\alpha
(\alpha -1)$. In addition, we will need to divide by the permutation
factor $2!$ if the two indices $i$ and $j$ are identical.  

In general, as we continue to the third and higher order terms,
the only things to keep in mind are the overall order of the term
(the value of the derivative of $f$ is fixed in this way), and the
instances of repeated indices where it will be necessary to divide 
by a multiplicity factor for each case where there is a repeated 
index.
For brevity, we represent the expansion series as 
$(1 + [\Delta_{1} + \Delta_{2} + \ldots + \Delta_{n}])^{\alpha} = 
1 + [A_{1} \Delta_{1} + A_{2} \Delta_{2} + \ldots + A_{n}
\Delta_{n}]
+ [A_{11} \Delta_{1}^{2} + A_{12} \Delta_{1} 
\Delta_{2} ] + \ldots$  
A few examples are
\begin{gather}
A_{12} = \alpha (\alpha - 1);~A_{11}
= \frac{1}{2!} \alpha (\alpha - 1) \\
A_{1222} = \frac{1}{3!} \alpha (\alpha -1)(\alpha -2)
(\alpha - 3)\\
A_{1234} = 
\alpha (\alpha -1) (\alpha -2)(\alpha -3) 
\end{gather}

Our results for the multivariable case also have relevance for perturbations
which are themselves power series in a single variable.
One may have, for example, an expression of the form
\begin{equation}
f(x) = (1 + [c_{1} x + c_{2} x^{2} + c_{3} x^{3} + \ldots])^{\alpha},
\label{eq:eq20}
\end{equation}
where the series may terminate or instead be an infinite power 
series.
To make use of the machinery we have discussed in the case of more 
than one variable, we substitute a $\Delta_{i}$ for each
term in the series (e.g. we set $\Delta_{1} \equiv c_{1} x$,
$\Delta_{2} = c_{2}x^{2}$, etc). For convenience in sorting the terms,
we choose the index $i$ to correspond to the degree of the term
it represents.  As an example, 
in developing the power series for $\sec (x)$, we 
begin with
\begin{equation}
\sec(x) = 1/[1 + (-x^{2}/2! + x^{4}/4! - x^{6}/6! + \ldots)],
\label{eq:eq21}
\end{equation}
and for practical purposes that will become apparent we 
follow the pattern mentioned above and use the substitution 
$\sec(x) = 1/(1 + \Delta_{2} + \Delta_{4} + \Delta_{6} + \ldots)$, 
where $\Delta_{2} = -x^{2}/2!$,
$\Delta_{4} = x^{4}/4!$, and so on.  

%To see how in this manner we build up the power series in $x$,
%let us examine the expression
%$f(x) = (1 + [\Delta_{1} + \Delta_{2} + \ldots] )^{\alpha}$
%\begin{align}
%\nonumber
%f(x) = 1 + A(y_{1}) \Delta_{1} + [A(y_{1}^{2}) \Delta_{1}^{2} + 
%A(y_{2}) \Delta_{1}^{2}] \\ \nonumber
%+ [A(y_{3}) \Delta_{3} + A(y_{2}\Delta_{1}) 
%\Delta_{2} \Delta_{1} + A(y_{1}^{3}) \Delta_{1}^{3}] \\ \nonumber
%+ [A(y_{4}) \Delta y_{4} + A(y_{3} y_{1}) \Delta_{3} \Delta_{1} 
%+ A(y_{2}^{2}) \Delta_{2}^{2}
%+ \\ \nonumber A(y_{2}y_{1}^{2}) \Delta_{2} \Delta_{1}^{2} 
%+ A(y_{1}^{4}) \Delta_{1}^{4}] +  
%[ A(y_{5}) \Delta_{5} + \\ \nonumber A(y_{4} y_{1}) \Delta_{4} \Delta_{1} + 
%A(y_{3} y_{2}) \Delta_{3} \Delta_{2} +  
%A(y_{3} y_{1}^{2}) \Delta_{3} \Delta_{1}^{2} + \\ \nonumber
%A(y_{2}^{2} y_{1}) \Delta_{2}^{2} \Delta_{1} +
%A(y_{2} y_{1}^{3}) \Delta_{2} \Delta_{1}^{3} +  
%A(y_{1}^{5}) \Delta_{1}^{5}] +
%\ldots
%\label{eq:eq22}
%\end{align}

With the formalism we have discussed, we may now systematically
and conveniently write down the terms in
the expansion for $[1 + (\Delta_{1}
+ \Delta_{2} + \ldots )]^{\alpha}$, though we mention one caveat;
it is important to list all possible combinations of the 
$\Delta_{i}$ factors in developing the expansion.  For terms up 
to fourth order, this is easy to do by inspection. For the 
higher order cases, one can systematically enumerate all of
the $\Delta_{i}$ combinations by listing all possible 
partitions of the integer corresponding to the order, and 
a recursive procedure for the partitioning is described in 
the Appendix.
%\begin{gather}
%\nonumber
%f(x) = 1 + A(\Delta_{1}) \Delta_{1} + [A(y_{2}) \Delta_{2} + 
%A(y_{1}^{2}) \Delta_{1}^{2} ] + \\ \nonumber
%[ A(y_{3}) \Delta y_{3} + A(y_{1} y_{2} ) \Delta y_{1} \Delta y_{2}
%+ A(y_{1}^{3}) \Delta y_{1}^{3} ] + \\ \nonumber [A(y_{4}) \Delta y_{4} + 
%A(y_{1} y_{3}) \Delta y_{1} \Delta y_{3} + 
%A(y_{2}^{2}) \Delta y_{2}^{2} \\ 
% + A(y_{2} y_{1}^{2}) \Delta y_{2} \Delta y_{1}^{2}
% + A(y_{1}^{4}) \Delta y_{1}^{4}] + \\ \nonumber
%   + \ldots
%[A(y_{4}) \Delta y_{4} + A(y_{1} y_{3}) \Delta y_{1} \Delta y_{3} +
%[A(y_{4}) \Delta y_{4} + A(y_{1} y_{3}) \Delta y_{1} \Delta y_{3} +
%A(y_{2}^{2}) \Delta y_{2}^{2} + A(y_{1}^{4}) \Delta y_{1}^{4} + \ldots
%\label{eq:eq23}
%\end{gather}
Since the value of the partial derivatives depends only on the total
number of differentiations, we will use the notation $d_{i}$ where, e.g.,
$d_{1} = \partial f /\partial y_{i}$ and $d_{2} =
\partial^{2}f/\partial y_{i}^{2} =
\partial^{2} f/\partial y_{i} \partial y_{j}$
Following the rules derived earlier, the coefficients of powers of 
$x$ are
\begin{equation} \nonumber
d_{1} \Delta_{1} = \alpha c_{1} x 
\end{equation}
for the first order contribution,
\begin{equation}
d_{1} \Delta_{2} + \frac{1}{2!}d_{2} \Delta_{1}^{2}  = 
\left[ \alpha c_{2} + \frac{1}{2!} \alpha (\alpha - 1) 
c_{1}^{2} \right]x^{2}
\end{equation}
for the second order piece,
\begin{gather}
d_{1} \Delta_{3} + d_{2} \Delta_{2} \Delta_{1} + 
\frac{1}{3!} d_{3} \Delta_{1}^{3}  = \\ \nonumber
\left [ \alpha c_{3} + \alpha (\alpha - 1) c_{2} c_{1} +
\frac{1}{3!} \alpha (\alpha -1) (\alpha -2) c_{1}^{3} \right]x^{3} 
\end{gather}
for the third order terms,
\begin{gather}
d_{1} \Delta_{4} + d_{2} \Delta_{3} \Delta_{1} + 
\frac{1}{2!} d_{2} \Delta_{2}^{2} + \\ \nonumber 
\frac{1}{2!} d_{3} \Delta_{2} \Delta_{1}^{2} +
\frac{1}{4!} d_{4} \Delta_{1}^{4} = \\
\Bigl [\alpha c_{4} + \alpha (\alpha -1) c_{3}c_{1} + 
\frac{1}{2!} \alpha (\alpha - 1) c_{2}^{2} 
 + \\  \nonumber 
\frac{1}{2!} \alpha (\alpha -1) (\alpha -2) c_{2} c_{1}^{2} + 
\frac{1}{4!} \alpha (\alpha -1) (\alpha -2) (\alpha -3) c_{1}^{4} 
\Bigr] x^{4} 
\\ \nonumber
\end{gather}
for the fourth order contribution. 
%begin{gather}
%d_{1} \Delta_{5} + d_{2} \Delta_{4} \Delta_{1} + 
%d_{2} \Delta_{3} \Delta_{2} + \\ \nonumber
%\frac{1}{2!} d_{3} \Delta_{3} \Delta_{1}^{2} +
%\frac{1}{2!} d_{3} \Delta_{2}^{2} \Delta_{1} + 
%\frac{1}{3!} d_{4} \Delta_{2} \Delta_{1}^{3} + \\ \nonumber
%\frac{1}{5!} d_{5} \Delta_{1}^{5} = 
%[ \alpha c_{5} + \alpha (\alpha -1) c_{4} c_{1} +
%\alpha (\alpha -1) c_{3} c_{2} + \\ \nonumber
%\frac{1}{2!} 
%\alpha (\alpha -1)(\alpha -2) c_{3}c_{1}^{2} + 
%\frac{1}{2!} \alpha (\alpha -1)(\alpha -2)
%c_{2}^{2} c_{1} + \\ \nonumber \frac{1}{3!} 
%\alpha (\alpha -1) (\alpha -2) (\alpha -3) c_{2} 
%c_{1}^{3} \\
% + \frac{1}{5!} 
%\alpha (\alpha -1)(\alpha -2)(\alpha -3)(\alpha -4) c_{1}^{5}
%] x^{5}
%\end{gather}
%for the fifth order piece.

\section{The two component case and a multipole expansion}

\subsection{Two component perturbations in general}

We find a particularly simple result for the case of a two
component perturbation, 
where in general we have $f(1 + [\Delta_{1} + 
\Delta_{2}])$ 
Since we will eventually replace $\Delta_{1} + 
\Delta_{2}$ with
$c_{1} x+ c_{2}x^{2}$ (i.e. as occurs often in generating functions 
for special functions such as the Legendre and Hermite 
polynomials), we will again separate the terms of the
expansion into groups where the sum of the subscripts is the same;
in this manner we will automatically have an appropriately
organized power series in $x$
when we make the substitution $y_{1} = c_{1}x$, $y_{2} = c_{2}x^{2}$.
As is discussed in the Appendix, the fact that 
in this context one partitions 
integers into 1's and 2's with no higher integers appearing 
greatly limits the number of terms which will 
be generated for each order, and the number of terms associated
with each order rises only linearly with the order $n$. 

With these ideas in mind, we see that the first order contribution is
$d_{1}\Delta_{1}$.  The second order piece, with two terms, is $(d_{2} 
\Delta_{2}
+ \frac{1}{2!}d_{2} \Delta_{1}^{2})$.  The third order component also
has two terms since there is no $\Delta_{3}$ term in the perturbation,
and we obtain $d_{2} \Delta_{1} \Delta_{2} + 
\frac{1}{3!} d_{3} \Delta_{1}^{3}$; likewise, there being 
no $\Delta_{4}$ term, the fourth order piece has the form
$d_{2} \Delta_{2}^{2} + \frac{1}{2!}d_{3} \Delta_{2} \Delta_{1}^{2} 
+ \frac{1}{4!} d_{4} \Delta_{1}^{4}$,
and one continues in this manner for the higher order terms.  
If we specialize to the case 
$y_{1} = c_{1} x$ and $y_{2} =c_{2} x^{2}$, we find for the power series 
in $x$ 
%\begin{gather}
%\nonumber    
%f(1 + c_{1}x + c_{2} x^{2}) = f(1) + d_{1}c_{1}x + \left ( d_{1}c_{2} 
%+ \frac{1}{2!} d_{2}c_{1}^{2} \right )x^{2}  \\ \nonumber
%+ \left( d_{2} c_{2} c_{1} + \frac{1}{3!}d_{3} c_{1}^{3} \right)
%x^{3}  \\ \nonumber + \left( \frac{1}{2!} 
%d_{2}c_{2}^{2} + \frac{1}{2!} d_{3} 
%c_{2} c_{1}^{2} + \frac{1}{4!} d_{4} c_{1}^{4} 
%\right) x^{4}+  \\ \nonumber  \left( \frac{1}{2!} d_{3}
%c_{2}^{2} c_{1} + \frac{1}{3!} d_{4} c_{2} c_{1}^{3} 
%+ \frac{1}{5!} d_{5} c_{1}^{5} \right) x^{5} + \ldots
%\nonumber
%\end{gather}
%We have
\begin{gather}
\nonumber
f(1 + c_{1}x + c_{2}x^{2}) = f(1) + \Bigl \{ \frac{1}{0!0!}
 d_{1} c_{2}^{0}c_{1}^{1}x + \\ \nonumber 
+ \left( \frac{1}{1!1!}
 d_{2} c_{2}^{1} c_{1}^{1}
+ \frac{1}{0!3!}  d_{3}
c_{2}^{0} c_{1}^{3} \right) x^{3}  \Bigr \} \nonumber  \\
+ \Bigl \{ \left ( \frac{1}{1!0!}  d_{1} c_{2}^{1}c_{1}^{0}
 + \frac{1}{0!2!} d_{2} c_{2}^{0} c_{1}^{2} 
\right) x^{2} \\
+ \left( \frac{1}{2!0!}  d_{2} c_{2}^{2}c_{1}^{0} + 
\frac{1}{1!2!}  d_{3} c_{2}^{1} c_{1}^{2} 
+ \frac{1}{0!4!} 
d_{4}c_{2}^{0} c_{1}^{4}  \right) x^{4} + \ldots \Bigr \}
\end{gather}
where $0!$, $1!$, $c_{2}^{0}$, and 
$c_{1}^{0}$ have been inserted strategically and odd and 
even terms separated
to emphasize general trends. 
%\begin{gather}
%\nonumber
%f(1 + c_{1} x + c_{2} x^{2} ) = \\ \nonumber 
%f(1) + \Bigl \{ \sum_{i=1}^{1} \frac{1}{(1-i)!(2i-1)!} d_{i}
%c_{2}^{1-i} c_{1}^{2i -1} x \\ \nonumber
%+ \sum_{i=1}^{2}
%\frac{1}{(2-i)!(2i-i)!} d_{1+i}
%c_{2}^{2-i} c_{1}^{2i-1} x^{3} + \ldots \Bigl\} \\ \nonumber
%+\Bigl \{ \sum_{i=0}^{1} \frac{1}{(1-i)!2i!}  
%d_{1+i}c_{2}^{1-i}c_{1}^{2i} x^{2} +\\ 
%\sum_{i=0}^{2}
%\frac{1}{(2-i)!2i!} d_{2+i} c_{2}^{2-i}c_{1}^{2i} x^{4} + \ldots \Bigl \}
%\end{gather}
To obtain expressions for general even and odd powers, we 
combine the terms in parentheses into summations and
thereby make the 
dependence on the order $n$ more transparent.  We deduce for
the coefficients of $x$ the expressions 
\begin{gather}
{\mathcal A_{2n}^{\mathrm{even}}}  = 
\sum_{i=0}^{n} \frac{1}{(n-i)!2i!} d_{n+i}c_{2}^{n-i}
c_{1}^{2i} \\
{\mathcal A_{2n-1}^{\mathrm{odd}}} = \sum_{i=1}^{n} \frac{1}{(n-i)!(2i-1)!} 
 d_{n+i-1} c_{2}^{n-i} c_{1}^{2i-1}
\end{gather}
where we write for the full expansion $f(1 + [c_{1}x + c_{2} x^{2}]) = 
f(1) + {\mathcal A}_{1} x + {\mathcal A}_{2}x^{2} + {\mathcal A}_{3}x^{3} 
+ \ldots$

\subsection{Multipole expansion for an electrostatic potential}

A salient example from electrostatics where we may apply this 
logic is the multipole 
expansion of the electric 
potential of a charge displaced from the origin by $\Delta z$ as shown in 
Figure~1.  If we operate in the polar system, the combined 
potential due to the negative and positive charges is
\begin{gather}
\nonumber
V(r,\theta) =  - 
 Q/r[(1 - 2  \cos \theta \Delta/r+ 
\Delta^{2}/r^{2} )^{-1/2} \\
- (1 + 2 \cos \theta \Delta/r
+ \Delta^{2}/r^{2} )^{-1/2}] 
\end{gather}
For distances large relative to the size of 
the charge arrangement (i.e. where $r \gg \Delta$), 
we may expand in the small parameter $\Delta/r$. 

The perturbation is a two term series in $\Delta/r$;  
following the procedure described in this report, we set
$(1 + \Delta_{1} + \Delta_{2} )^{-1/2} \equiv
(1 - 2 \Delta /r  \cos \theta +  \Delta^{2}/r^{2} )^{-1/2}$
for the first term in the electrostatic potential.  The 
analysis of the second term, except for a sign 
difference, proceeds in an identical way.
%Expanding, we obtain
%\begin{gather}
%\nonumber
%(1 + \Delta_{1} + \Delta y_{2})^{-1/2} = 
%1 + A(y_{1}) \Delta y_{1} \nonumber\\
%+ A(y_{2}) \Delta y_{2} + 
%A(y_{1}^{2}) \Delta y_{1}^{2} + A(\Delta y_{1} \Delta y_{2} ) 
%\Delta y_{1} \Delta y_{2} \\ 
%= 1 + (-1/2)\Delta y_{1} + (-1/2) \Delta y_{2} + \frac{1}{2!}
%\frac{-1}{2} \frac{-3}{2} \Delta y_{1}^{2} + 
%\end{gather}
Setting $c_{1} = -2 \cos \theta$,
$c_{2} = 1$, and ``$x$''$=\Delta/r$,
we see that the first order contribution will be 
\begin{equation}
d_{1} c_{1} x = -(1/2)( -2 )\cos \theta (\Delta/r) = \cos \theta (\Delta/r),
\end{equation}
and this is $P_{1} (\cos \theta) (\Delta/r)$ where the
$P_{l}(x)$ are Legendre polynomials of order $l$.
The second order piece is 
\begin{gather}
\nonumber
\left( d_{1} c_{2} +\frac{1}{2!} d_{2} c_{1}^{2}
\right) x^{2} = \\ \nonumber
[ -1/2 + (1/2)(-1/2)(-3/2)4 \cos^{2} \theta] (\Delta/r)^{2}  \\
= (-1/2 + 3/2 \cos^{2} \theta )(\Delta/r)^{2} = P_{2} (\cos \theta ) 
(\Delta/r)^{2}
\end{gather}
The third order contribution is 
\begin{gather}
\nonumber
\left(  d_{2} c_{1} c_{2} + \frac{1}{3!} d_{3} c_{1}^{3} \right) = 
\\ \nonumber
[(-1/2)(-3/2)(-2 \cos \theta) + \\ \nonumber (1/6)(-1/2)(-3/2)(-5/2)
(-8 \cos^{3} \theta) ](\Delta/r)^{3} =\\
(-3/2 \cos \theta + 5/2 \cos^{3} \theta )(\Delta/r)^{3} = 
P_{3} (\cos \theta ) (\Delta/r)^{3}
\end{gather}
The fourth and higher order terms are obtained in a similar manner.
%with the fourth order case given by
%\begin{gather}
%\nonumber
%\left( \frac{1}{2!} d_{2} c_{2}^{2} + \frac{1}{2!} 
%d_{3} c_{2} c_{1}^{2}  + \frac{1}{4!} c_{1}^{4} \right) = \\
%(3/8 -15/4\cos^{2} \theta + 35/8 \cos^{4} \theta) (\Delta/r)^{4},
%\end{gather}
%which we recognize as $P_{4} (\cos \theta ) (\Delta/r)^{4}$,
%and the fifth order piece being
%\begin{gather}
%\nonumber
%\left( \frac{1}{2!} d_{3} c_{2}^{2} c_{1} + \frac{1}{3!} d_{4} 
%c_{2} c_{1}^{3} + \frac{1}{5!} d_{5} c_{1}^{5} \right) = \\
%(15/8 \cos \theta - 35/4 \cos^{3} \theta + 63/8 \cos^{5} \theta ),
%\end{gather}
%identical to $P_{5} (\cos \theta ) (\Delta/r)^{5}$
We have for the combined potential 
\begin{gather}
V(r, \theta) = \frac{2 \Delta Q}{r^{2}} \cos \theta + 
\frac{ \Delta^{3} Q}{r^{4}} ( -3 \cos \theta + 5 \cos^{3} \theta) + \ldots
\end{gather}
We recognize the first nonzero term as the dipolar contribution where 
the factor $2 \Delta Q$ is
precisely the dipole moment appropriate to the geometry
and the equal and opposite charges $+Q$ and $-Q$.

\subsection{Additional Examples:  Hermite Polynomials and 
the Bernoulli numbers}
Generating functions often appear in the form $f[1 + (c_{1}x + c_{2}x^{2})]$,
and we may therefore apply the formalism used in the context of the 
electric multipole expansion to generate the first several 
polynomials in such cases.  For the Hermite polynomials, the generating function
$e^{tx - t^{2}/2} = g(x,t)$~\cite{dos} is particularly amenable to 
our expansion technique since 
the derivatives $d_{1} = d_{2} = \ldots = d_{n} = 1$ are each 
equal to unity. With $e^{tx - t^{2}/2} = \sum_{i = 0}^{\infty}
a_{n} t^{n}$ with $H_{n}(x) = n! a_{n}$, we find that  
$H_{0}(x) = 1$ and $H_{1}(x) = (1!)(c_{1}) = x$, while 
$H_{2}(x)$
and $H_{3} (x)$ are given by
\begin{gather}
H_{2} (x) = 2! \left (c_{2} + \frac{1}{2}c_{1}^{2} \right) = 
2 \left (-\frac{1}{2} + \frac{x^{2}}{2} \right ) = x^{2} - 1, 
\end{gather} 
and
\begin{gather}
H_{3} (x) = 3 !\left( c_{2} c_{1} 
+ \frac{1}{3!} c_{1}^{3} \right ) = 6 \left (-\frac{x}{2}  
+ \frac{x^{3}}{6} \right ) = x^{3} - 3x.
\end{gather}

Our technique can be used to conveniently 
obtain the Bernoulli numbers $B_{n}$ 
from~\cite{tres}
\begin{gather}
\frac{x}{e^{x} -1} = 1 + B_{1} x + \frac{B_{2}}{2!} x^{2} + \dots 
\end{gather}
It is useful to write 
\begin{gather}
\frac{x}{e^{x} - 1} = \frac{1}{(1 + x/2! + x^{2}/3! + \ldots)}
\end{gather}
We see that $B_{1} = d_{1} c_{1} = -\frac{1}{2}$. For $B_{2}$, we
have
\begin{gather}
B_{2} = 2! \left ( d_{1}c_{2} + \frac{1}{2!} d_{2} c_{1}^{2} \right) = 
2 ( -1/6 + 1/4) = \frac{1}{6}
\end{gather} 
We find for $B_{3}$ 
\begin{gather}
\nonumber
B_{3} = 3!\left( d_{1}c_{3} + d_{2} c_{1} c_{2} + \frac{1}{3!}
d_{3} c_{1}^{3} \right) \\ = 6(-1/24 + 1/6 + 1/8) = 0
\end{gather}

\subsection{Conclusions}
In conclusion, we have discussed a technique for facilitating the 
manipulation of power series which relies on the structure of the
multivariable translation operator when written as an infinite 
product.  The technique eliminates intermediate calculations 
while minimizing clutter.  The approach has been demonstrated for series 
raised to powers, but the technique may be applied in 
more general situations 
where one seeks to expand a function of the form $f[1 + (a_{1}x + 
a_{2}x^{2} + \ldots)]$  
where the perturbing influence is itself a 
power series in a small quantity $x$.  In the case of a two 
component perturbation (e.g. as in generating functions for special
functions such as the Legendre and Hermite polynomials),
the process of generating the power 
series is relatively simple, and our technique facilitates the 
derivation of a analytical closed form expression for the 
coefficients of each order of $x$.

\section{Appendix:  The Partitioning of integers}

The task of exhausting all possible combinations of $\Delta y_{i}$ 
becomes increasingly subtle with increasing order. As an example,
for the second order case one has only $\left \{ \Delta_{2}, 
\Delta_{1}^{2} \right \}$, whereas for 
fifth order the set of possibilities is much larger, namely 
$\left \{ \Delta_{5}, \Delta_{4} \Delta_{1}, 
\Delta_{3} \Delta_{2}, \Delta_{3} \Delta_{1}^{2}, 
\Delta_{2}^{2} \Delta_{1}, \Delta_{2} \Delta_{1}^{3}, 
\Delta_{1}^{5} \right \} $.
Nevertheless, one can easily list all possibilities for small to moderate
order by following a recursive partitioning procedure which we discuss
here.  We first examine the general case where the perturbing 
influence may have a large (or even infinite) number of terms.
We then find a much simpler result for the case of a two component
perturbation.

\subsection{The general case}

Let us denote the set of all possible partitions
 of the integer $n$ with curly braces, (i.e. as $\left \{ n \right \}$).
It is easy to see that for $n = 1$, there is only
$\left \{ 1 \right \}  = [1]$. 

Next, 
we discuss a systematic procedure for generating partitions by
examining successively higher integers beginning with $n = 2$ and 
continuing through $n = 6$.
One always begins with the unary partitioning $[2]$, just the number 
itself enclosed in square brackets. The next step 
is to generate all unique binary
partitions, and for $n = 2$ there is one of these, $[1,1]$. To represent the 
set of possible partitions, we use the group addition 
symbol $\oplus$ to indicate the combination of
the individual partitions.  E.G., for $n = 2$, we have
\begin{equation}
\left \{  2 \right \} = [2] \oplus [1,1]
\end{equation} 

For $n = 3$, we proceed in a similar way, again, 
forming all possible binary partitions and finding
$\left \{  3 \right \} = [3] \oplus [ \left \{ 2 \right \} , 1]$.
Now, substituting the partitioning $ \left \{ 2 \right \}$ determined
for $n = 2$, we obtain
\begin{equation}
\left \{ 3 \right \} = [3] \oplus [[2] \oplus [1,1],1]
\end{equation}
Exploiting linearity and distributing yields
\begin{equation}
\left \{ 3 \right \} = [3] \oplus [2,1] \oplus [1,1,1]
\end{equation}
for the partitioning corresponding to $n = 3$.  

For $n = 4$, we have 
\begin{equation}
\left \{ 4 \right\} = [4] \oplus [\left \{ 3 \right \} ,1] 
\oplus [\left \{ 2 \right \}, \left \{ 2 \right \} ]
\end{equation}
Acting recursively and inserting the relations we obtained
for $\left \{3 \right \}$ and $\left \{ 2 \right \}$ yields
\begin{gather}
\nonumber
\left \{ 4 \right \} = [4] \oplus [ [3] \oplus [2,1] \oplus [1,1,1],1]\\
\oplus [ [2] \oplus [1,1],[2] \oplus [1,1]]
\end{gather}
Distributing and eliminating redundant terms yields
\begin{equation}
\left \{ 4 \right \} = [4] \oplus [3,1] \oplus [2,2] \oplus [2,1,1] 
\oplus [1,1,1,1]
\end{equation}
We may proceed in the same manner for $n = 5$ and higher, but the
structure of the partitionings $\left \{ 2 \right \}$, 
$\left \{ 3 \right \}$, and $\left \{ 4 \right \}$ suggests a 
more efficient and straightforward approach akin to counting.
To emphasize the counting character of the partitioning process,
we have taken care to rank the specific partitions in decreasing 
``dictionary'' order, where the first entry in the partitioning
determines priority and ties are broken by subsequent entries 
where needed.  We illustrate the systematic counting process
for the case $n = 6$, and we obtain $\left \{ 6 \right \}$ in this 
manner with no discarded terms.

The highest possible number which may appear in $\left \{6 
\right \}$ is just a single ``6'', so the first entry in the 
partitioning is $[6]$.  The next highest entry would be a ``5'',
and the only possible partition is $[5,1]$.  The highest ranking 
entry in the next tier is $[4,2]$, and we may decrement by breaking up
the lowest reducible entry;  we hence split the 
``2'' into two ``1'''s yielding $[4,1,1]$.  For the next level, the 
highest ranking entry is $[3,3]$. The only way to generate
a lower entry is to reduce the second  ``3'', yielding 
$[3,2,1]$, followed by $[3,1,1,1]$. The next level 
and the final tier 
will consist of the entries $[2,2,2] \oplus [2,2,1,1] \oplus
[2,1,1,1,1] \oplus [1,1,1,1,1,1]$. To further illustrate the 
decrementation method, we generate the partitioning for $n=8$ 
by counting down from $[8]$.
\begin{gather}
\nonumber
\left \{ 8 \right \} = [8] \oplus [7,1] \\ \nonumber \oplus [6,2] 
\oplus [6,1,1] \\ \nonumber \oplus [5,3] \oplus [5,2,1] \oplus [5,1,1,1] 
\\ \nonumber \oplus [4,4] \oplus [4,3,1]  \oplus [4,2,2] \oplus [4,2,1,1] 
\oplus [4,1,1,1,1] \\ \nonumber \oplus [3,3,2]  
\oplus [3,2,2,1] \oplus 
[3,2,1,1,1] \oplus [3,1,1,1,1,1]  \\ \nonumber \oplus [2,2,2,2] \oplus
[2,2,2,1,1] \oplus [2,2,1,1,1,1] \oplus [2,1,1,1,1,1,1] \\
\oplus [1,1,1,1,1,1,1,1]
\end{gather}  

\subsection{The case of a two component perturbation}
With a two component perturbation such as $\Delta_{1} 
+ \Delta_{2} = c_{1}x + c_{2}x^{2}$, the process of partitioning the
integers for each order in $x$ simplifies considerably, and can 
be done essentially by inspection for any order.  As an example, 
we have for the first several orders
\begin{gather}
\nonumber
\left \{ 1 \right \} = [1];~~~ \left \{ 2 
\right \} = [2] \oplus [1,1] \\ \nonumber
\left \{ 3 \right \} = [2,1] \oplus [1,1,1];~~~
\left \{ 4 \right \} = [2,2] \oplus [2,1,1]
\oplus [1,1,1,1] \\ \nonumber
\end{gather}

\begin{figure}
\label{Fig:fig1}
\centerline{\includegraphics[width=2.2in]{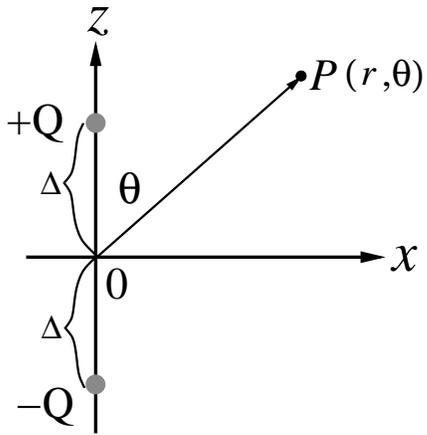}}
\caption{The charge arrangement with $Q$ and $-Q$ on the $z$ axis and 
separated by a distance $2 \Delta$.}
\end{figure}

\end{document}